\documentclass[]{aa}

\usepackage{graphicx}
\usepackage{txfonts}
\usepackage[svgnames]{xcolor}
\usepackage[colorlinks=true, linkcolor = blue, urlcolor = magenta, hyperfootnotes=false]{hyperref}
\usepackage{multirow}
\usepackage{footmisc}
\newcommand{\astfootnote}[1]{%
\let\oldthefootnote=\thefootnote%
\setcounter{footnote}{0}%
\renewcommand{\thefootnote}{\fnsymbol{footnote}}%
\footnote{#1}%
\let\thefootnote=\oldthefootnote%
}
\usepackage{xcolor}
\usepackage{makecell}

\usepackage{tablefootnote}

\hypersetup{
    colorlinks,
    citecolor=blue
}

\usepackage{natbib}
\bibpunct{(}{)}{;}{a}{}{,} % to follow the A&A style
\usepackage{subfigure}
\usepackage{wasysym}
\usepackage{siunitx}

\begin{document} 

   \title{Tentative co-orbital submillimeter emission within the Lagrangian region $L_5$ of the protoplanet PDS~70~b}

   \author{O.~Balsalobre-Ruza\inst{\ref{cab}}
         \and
          I.\,de\,Gregorio-Monsalvo\inst{\ref{eso}}
          \and
          J.~Lillo-Box
          \inst{\ref{cab}}
          \and
          N.~Huélamo\inst{\ref{cab}}
          \and
          Á.~Ribas\inst{\ref{cambridge}}
          \and
          M.~Benisty\inst{\ref{nice},\ref{gre}}
          \and
          J.~Bae\inst{\ref{flo}}
          \and
          S.~Facchini\inst{\ref{mil}}
          \and
          {R.~Teague\inst{\ref{usa1}}
          }}

   \institute{Centro de Astrobiología (CAB), CSIC-INTA, Camino Bajo del Castillo s/n, 28692, Villanueva de la Cañada, Madrid, Spain\\
   e-mail: {\tt obalsalobre@cab.inta-csic.es}\label{cab}
   \and
    European Southern Observatory, Alonso de Córdova 3107, Casilla 19, Vitacura, Santiago, Chile\label{eso}
    \and
    Institute of Astronomy, University of Cambridge, Madingley Road, Cambridge, CB3 0HA, UK\label{cambridge}
    \and
    Laboratoire Lagrange, Université Côte d’Azur, CNRS, Observatoire de la Côte d’Azur, 06304 Nice, France\label{nice}
    \and
    Univ. Grenoble Alpes, CNRS, IPAG, 38000 Grenoble, France\label{gre}
    \and
    Department of Astronomy, University of Florida, Gainesville, FL 32611, USA\label{flo}
    \and
    Dipartimento di Fisica, Universit\`a degli Studi di Milano, Via Celoria 16, I-20133 Milano, Italy\label{mil}
    \and
    Department of Earth, Atmospheric, and Planetary Sciences, Massachusetts Institute of Technology, Cambridge, MA 02139, USA\label{usa1}}
    
   \date{Received ; accepted }

% \abstract{}{}{}{}{} 
% 5 {} token are mandatory
  \abstract
   {High-spatial resolution Atacama Large Millimeter/submillimeter Array (ALMA) data have revealed a plethora of substructures in protoplanetary disks. Some of those features are thought to trace the formation of embedded planets. One example is the gas and dust that accumulated in the co-orbital Lagrangian regions $L_4$/$L_5$, which were tentatively detected in recent years and might be the pristine material for the formation of Trojan bodies.}
   {This work is part of the TROY project, whose ultimate goal is to find robust evidence of exotrojan bodies and study their implications in the exoplanet field. Here, we focus on the early stages of the formation of these bodies by inspecting the iconic system PDS~70, the only confirmed planetary system in formation.} 
   {We reanalyzed archival high-angular resolution Band 7 ALMA observations from PDS~70 by doing an independent imaging process to look for emission in the Lagrangian regions of the two detected gas giant protoplanets, PDS~70~b and c. We then projected the orbital paths and visually inspected emission features at the regions around the $L_4$/$L_5$ locations as defined by $\pm$ 60$^{\circ}$ in azimuth from the planet position.}
   {We found emission at a $\sim$4-$\sigma$ level ($\sim$6-$\sigma$ when correcting from a cleaning effect) at the position of the $L_{5}$ region of PDS~70~b. This emission corresponds to a dust mass in a range of 0.03\,--\,2\,M$_{Moon}$, which potentially accumulated in this gravitational well.}
   {The tentative detection of the co-orbital dust trap that we report requires additional observations to be confirmed. We predict that we could detect the co-orbital motion of PDS~70~b and the dust presumably associated with $L_5$ by observing again with the same sensitivity and angular resolution as early as February 2026.}


   \titlerunning{\texttt{}}
   \maketitle
%
%-------------------------------------------------------------------

\section{Introduction}

Trojan asteroids are common inhabitants of the Solar System. They are minor bodies\footnote{The largest known is the Jupiter Trojan (624)~Hektor, a bilobe-shaped body with an equivalent diameter of $\sim$\,220\,km (\citealt{2014ApJ...783L..37M}).} populating the $L_4$ and $L_5$ Lagrange regions of a planet, leading and trailing it 60$^{\circ}$ apart in the same orbital path. \citet{2002AJ....124..592L} theoretically demonstrated that Trojans as massive as the main planet could be long-term stable, thus inspiring the concept of co-orbital planets. The fact that the stability condition for co-orbitals is met for pairs of bodies with similar masses opens the possibility of searching for exotrojans using the same methods and instruments as for the currently confirmed exoplanets (e.g., radial velocities, \citealt{2012MNRAS.421..356G}, \citealt{2015A&A...581A.128L}; transit timing variations, \citealt{2013CeMDA.117...75H}; transits themselves, \citealt{2015ApJ...811....1H}; or the combination of transit and radial velocity data, \citealt{2017A&A...599L...7L}). However, the efforts in the search for exotrojan planets have resulted in a select number of unconfirmed candidates so far (\citealt{2014A&A...562A.109L}; \citealt{2015ApJ...811....1H}; \citealt{2018A&A...609A..96L}; \citealt{2018A&A...618A..42L}).\vspace{0.2cm}

Several numerical and hydrodynamical simulations on the evolution of planets embedded in protoplanetary disks agree in converging to solutions compatible with Trojan formation. These simulations show that dust particles preferentially accumulate at the $L_4$ and $L_5$ regions of the protoplanet (e.g., \citealt{2002AJ....124..592L}; \citealt{2007A&A...463..359B}; \citealt{2020A&A...642A.224M}).  
 The enhanced dust growth at these locations can form planetesimals and, eventually, rocky bodies with masses as high as those of super-Earths (e.g., \citealt{2009A&A...493.1125L}). Therefore, Trojans might be a natural byproduct of planetary formation through in situ formation mechanisms. Their assembling could be studied at the earliest stages, when protoplanets are still embedded within protoplanetary disks. Furthermore, they might be unique targets to study what protoplanet interiors are made of since Trojans and planets are most likely formed simultaneously (e.g., \citealt{2019ApJ...884L..41B} results suggest that grains have difficulties accumulating in the Lagrangian regions once the planet has carved a gap). \vspace{0.2cm}

High-angular resolution observations of protoplanetary disks (mainly in the submillimeter range) have shown that the presence of substructures such as gaps and rings is very common (e.g., \citealt{2018A&A...620A..94G}; \citealt{2020ARA&A..58..483A}; \citealt{2022arXiv220309991B}). In particular, two of those disks show continuum emissions that could be explained by dust accumulations in the Lagrangian regions of nondetected planets, thus becoming an indirect hint for the presence of such young protoplanets. One case is \object{HD~163296}, which shows a well-defined arc-like feature inside a gap satisfactorily reproduced by simulations of dust corotating with a potential protoplanet (\citealt{2018ApJ...869L..49I}; \citealt{2021A&A...647A.174R}; \citealt{2023arXiv230113260G}). The other is \object{LkCa~15}, for which \citealt{2022ApJ...937L...1L} have reported two emissions at $\sim$10-$\sigma$ significance separated by 120$^\circ$ in azimuth with shapes comparable to those found in the simulations (e.g., \citealt{2018ApJ...869L..47Z}). These claims are an indirect method to infer the presence of (as of yet undetected) protoplanets, and also the first piece of evidence of exotrojan formation. \vspace{0.2cm}

The K7 T-Tauri star \object{PDS~70} (\citealt{2016MNRAS.461..794P}; \citealt{2018A&A...617L...2M}) is a unique target for planetary formation studies since it harbors the only robust detection of two protoplanets. It is surrounded by a highly structured protoplanetary disk with a wide inner cavity presumably carved by the planets. Both planets have been observed at infrared and submillimeter wavelengths and at different epochs (e.g., \citealt{2018A&A...617A..44K}, \citealt{2018ApJ...863L...8W}, \citealt{2019NatAs...3..749H}, \citet{2019ApJ...879L..25I}, \citealt{2021ApJ...916L...2B}). This has enabled their orbits to be modeled  and they appear to be migrating into a 2:1 mean motion resonance (\citealt{2019ApJ...884L..41B}; \citealt{2021AJ....161..148W}). As an example of how insightful this system is for planet formation studies, \citealt{2021ApJ...916L...2B} recently reported the detection of emission colocated with \object{PDS~70~c}, which possibly is the first detection of a circumplanetary disk (CPD). \vspace{0.2cm}

For this paper, through a reanalysis of archival public Atacama Large Millimeter/submillimeter Array (ALMA) data, we searched for excess emission compatible with dust accumulation in the Lagrangian regions of both protoplanets around \object{PDS~70}. This work is part of the TROY\footnote{\url{https://www.troy-project.com/}} project (\citealt{2018A&A...609A..96L}), which is devoted to searching for the first exotrojans and studying their impact in planetary systems. In Sect.~\ref{sec:obs} we describe the observational data, and we present the results of our search in Sect.~\ref{sec:results}. Section~\ref{sec:dis} is dedicated to discussing a potential detection of dust accumulation in the $L_5$ region of \object{PDS~70~b}. We provide our conclusions in Sect.~\ref{sec:concl}.

\begin{figure*}
  \begin{center}
     \subfigure{\includegraphics[width=185mm]{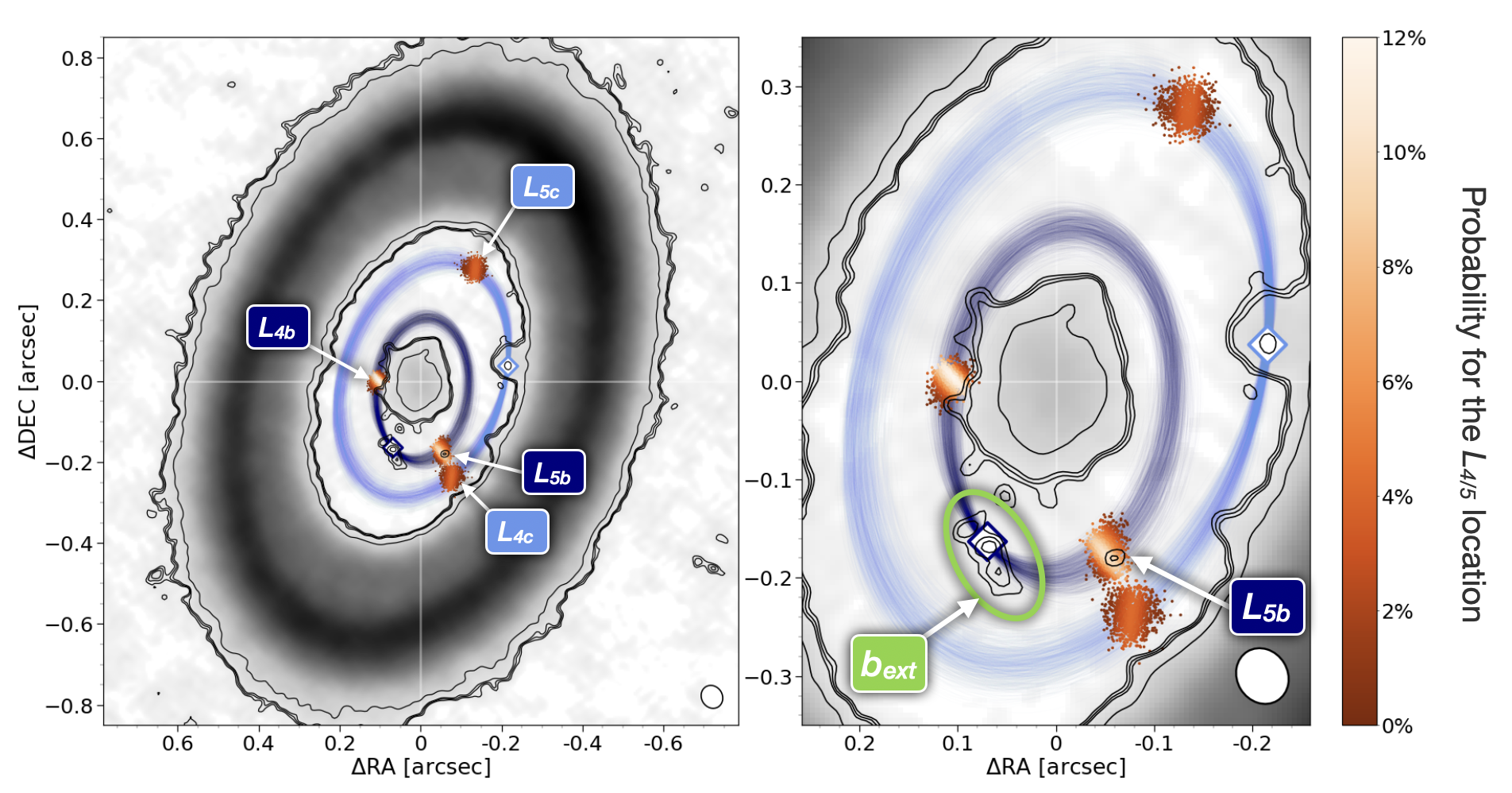}}
  \end{center}
 \caption{PDS\,70 ALMA observations with the projected distributions of the projected planetary orbits and Lagrangian points ($L_4$ and $L_5$) distributions. The background image (inverted grayscale) corresponds to the continuum observations from the combined ALMA data (see Section \ref{sec:obs}). The locations for the corresponding $L_4/L_5$ points for each orbit (total of 10$^3$ for each planet) are plotted with dots, whose density is represented by the colors from the right-hand side bar (it is important to note that lighter colors denote a higher density of probability for the location). The origin of coordinates is shown with white lines. The synthesized beam (0.058\arcsec\,$\times$\,0.052\arcsec, PA\,=\,58.71\,$^{\circ}$) is represented in the bottom right of each panel by a white ellipse. Diamond-shaped markers indicate the planets locations. Contours correspond to 3, 3.5, 4, and 8-$\sigma$. \textit{Left:} Global picture of PDS\,70. \textit{Right:} Zoom of the inner cavity to inspect the $L_5$ region of PDS~70~b. In green the extended emission associated with planet b reported by \citealt{2019ApJ...879L..25I} is indicated.} 
 \label{fig:projection}
\end{figure*}

\section{Observations}
\label{sec:obs}

In this work, as a starting point, we use the combined self-calibrated ALMA dataset used by \citet{2021ApJ...916L...2B}. They included submillimeter continuum data in Band 7 at wavelength 855~$\mu$m from program IDs 2018.A.00030.S, 2015.1.00888.S, and 2017.A.00006.S. In particular, we worked with their combination of three observations gathered at different epochs and different baselines. For more details on the calibration and the centering of these datasets, we refer the reader to the seminal publication.\vspace{0.2cm}

The difference between the combined image shown in \citet{2021ApJ...916L...2B} and the one from this work is the imaging procedure. We used the Common Astronomy Software Applications package (CASA; \citealt{2007ASPC..376..127M}) version 6.1.1. We imaged the visibilities of their self-calibrated data using the task \texttt{tclean}, a multifrequency synthesis mode, and a multiscale clean deconvolution algorithm with scales of 0, 1, and 2 times the beam full width at half maximum. Briggs weighting (\citealt{1992ASPC...25..170B}) was tested using different robust parameters; finally, we chose r~=~1.7 since it provided the best trade-off between sensitivity and angular resolution with the best signal-to-noise ratio (see Appendix \ref{sec:set_rob}). A map size of 2000~$\times$~2000 pixels was produced, with a pixel size of 0.005\arcsec. The final image was corrected from the primary beam response showing a synthesized beam size of 0.058\arcsec~$\times$~0.052\arcsec\,, a position angle of 58.71$^\circ$, and a root mean square (rms) equal to 11\,$\mu$Jy\,beam$^{-1}$ calculated as the standard deviation in the whole image not corrected for the primary beam but excluding the disk. In order to follow a very conservative analysis of the data, we performed our analysis without applying the correction that was used in \citet{2021ApJ...916L...2B} to deal with a cleaning effect first discribed in \citet{1995AJ....110.2037J}, the so-called JvM effect as the acronym of their names. Yet, we also show in Appendix \ref{sec:set_rob} the impact of the robust parameter for JvM-corrected images. For an explanation of this correction, readers can refer to Appendix \ref{sec:jvm} and \citet{2021ApJS..257....2C}.

\section{Results}
\label{sec:results}

The center of the combined ALMA image in the original publication (\citealt{2021ApJ...916L...2B}) is defined by the center of an ellipse fitting the emission maximum of the outer ring in the image plane. In order to avoid any effect induced by different cleaning processes, we centered the orbit using as a criterion that the reported position by the authors of planet c must match the maximum of its submillimeter emission. This corresponds with an offset with respect to the center of the observations of $\Delta$RA\,=\,9.5\,mas and $\Delta$DEC\,=\,12.5\,mas (center marked with white lines in Fig.\,\ref{fig:projection}), which is compatible with their uncertainties and also is in good agreement with the offset reported by the authors when modeling the outer disk with the \texttt{frank} (\citealt{2020MNRAS.495.3209J}) package ($\Delta$RA\,=\,12\,mas and $\Delta$DEC\,=\,15\,mas). \vspace{0.2cm}

We projected the orbits of \object{PDS~70~b} and c onto the image based on the orbital parameters derived by \citet{2021AJ....161..148W}. We generated 10$^3$ random orbits for each planet taking normal distributions centered in the dynamically stable parameters within 1-$\sigma$ (see Table\,3 from their work). Additionally, we restricted the orbits to those that cross the positions of the submillimeter peak emissions associated with both planets in our image (as reported by \citealt{2021ApJ...916L...2B}). For each of the orbits, we located the corresponding minimum of the gravitational well of the $L_4$ and $L_5$ points, at $\pm$\,60$^{\circ}$ in azimuth from those peak intensities within the orbits. In Fig.~\ref{fig:projection} we show the PDS~70 continuum image with the contour lines corresponding to 3, 3.5, 4, and 8-$\sigma$, where $\sigma$ is the rms of the image as stated in Sect. \ref{sec:obs}. The Table\,\ref{tab:L4L5app} from Appendix \ref{sec:astrometry} lists the positions for the maximum of the submillimeter emission associated with the planets and their Lagrangian points at the time of the long baseline observations (July 2019). \vspace{0.2cm} 

From Fig.\,\ref{fig:projection}, we visually find a match between a submillimeter unresolved compact emission detected with a significance of near 4-$\sigma$ ($\sim$6-$\sigma$ with the JvM-corrected image, see Appendix~\ref{sec:jvm}) and the expected position of the $L_5$ point of the submillimeter extended emission associated with \object{PDS~70~b} (hereafter $L_{5b}$\footnote{We use equivalent notation for the other Lagrangian regions.}). Conversely to planet c, the peak of the submillimeter emission associated with PDS~70\,b is shifted from the infrared centroid as already discussed by \citet{2019ApJ...879L..25I} and \citet{2021ApJ...916L...2B}. In Appendix \ref{sec:bshift} we show the same exercise but considering the infrared position of the planet, concluding that this new compact emission is fully compatible with dust librating within the $L_{5b}$ region. \vspace{0.2cm}

We found this new emission using the same dataset published in \citet{2021ApJ...916L...2B} as the result of carrying out a cleaning focused on maximizing the signal-to-noise ratio within the disk cavity. Their goal was the analysis of the CPD around PDS~70\,c whose emission is noticeably stronger and thus they did not require a cleaning so close to the noise level. To reach the required sensitivity for this tentative detection, it has been key to use the combined data from the three different epochs. We have checked that over these epochs no effect is expected due to the motion of the emission in $L_{5b}$. Between August 2016 and July 2019, the motion following the orbit of planet b should have been around half of the beam size, and thus not being sensitive to it.\vspace{0.2cm}

We can carry out a rough estimation of the dust mass potentially trapped in the $L_{5b}$ region by using the equation derived by \citet{1983QJRAS..24..267H}, for which the dust mass ($M_d$) scales with the flux density ($F_{\nu}$):
\begin{equation}
    M_d = \frac{F_{\nu}d^2}{\kappa_{\nu}B_{\nu}(T)},
\end{equation}
where $d$ is the distance to the source, $\kappa_{\nu}$ the dust opacity, and $B_{\nu}(T)$ the blackbody emission for a source with an effective temperature $T$. We assume that the emission at 855\,$\mu$m is optically thin and that it comes from dust thermal emission. The estimated flux density of the compact emission, which is the same as the peak intensity for an unresolved source, is 40\,$\mu$Jy. We assume a dust temperature of 19\,K as estimated by \citet{2019A&A...625A.118K} due to the stellar irradiation at the separation of PDS~70~b ($\sim$\,22\,au), and we used the dust opacity from \citet{2018ApJ...869L..45B} at a close frequency as an approximation ($\kappa_{0.88\,mm}$\,=\,3.6\,cm$^2$\,g$^{-1}$). We consider an uncertainty in the temperature of $\pm$\,5\,K, a 25\% in the dust opacity, and $\sqrt{\mathrm{rms}^2 + \Delta F_{calib}^2}$  for the density flux, where we took $\Delta F_{calib}$ = 10\% of the measured density flux. Thus, the dust mass traced at this wavelength is in the range of 0.0004\,--\,0.02\,M$_{\oplus}$, or equivalently 0.03~--~2~M$_{Moon}$.

\begin{figure*}
  \begin{center}
     \subfigure{\includegraphics[width=180mm]{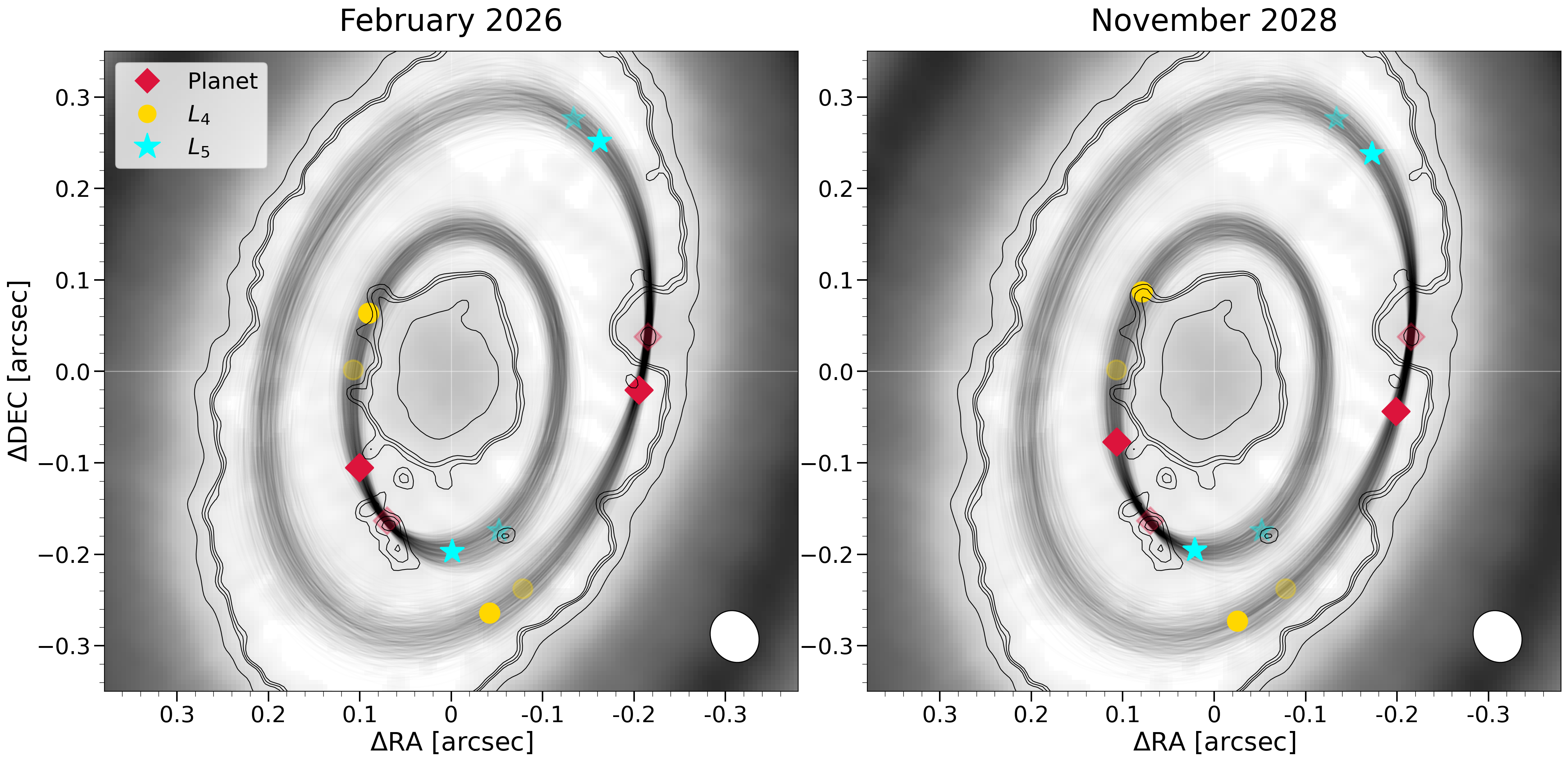}}
  \end{center}
 \caption{Prediction for the positions of the planets and their Lagrangian points at different epochs. \textit{Left:} Minimum observing date to detect the co-orbital motion in the orbit of planet PDS\,70~b. \textit{Right}: Equivalent for the orbit of planet PDS~70~c. Transparent symbols are the position in the current image epoch (July 2019), while solid symbols are for the epoch in the title. Same ALMA continuum image with the same projected orbits as in Fig.\,\ref{fig:projection}. Contours are 3, 3.5, 4, and 8-$\sigma$.}
 \label{fig:mov}
\end{figure*}

\section{Discussion}
\label{sec:dis}

Detecting the co-orbital motion of both the planet and the $L_{5b}$ emission along the expected orbital path would provide additional evidence in favor of the exotrojan dust scenario. We consider a minimum spatial separation equal to the synthesized beam to be able to detect motions of planets and Trojans between two epochs, considering that the size of the synthesized beam is larger than the absolute astrometric accuracy, which in a very unfavorable scenario of poor atmospheric phase stability could be $\sim$\,0.02\arcsec\, for a 4-$\sigma$ detection. Assuming that additional data of a similar quality (at least the same sensitivity and spatial resolution) are obtained in the near future, we studied the epoch when the orbital motion will be enough to move to a projected distance equivalent to the beam size. In Table\,\ref{tab:L4L5app} and Fig.\,\ref{fig:mov}, the July 2019 positions of the planets and their Lagrangian regions with their expected locations in February 2026 and November 2028 are compared. In the case of PDS~70~b, it will have moved a projected distance of 0.065\arcsec\, by February 2026. This is the earliest that we will be able to detect the motion of the three relevant locations of the orbit (planet b, $L_{4b}$, and $L_{5b}$). This date is 6.6\,years after the epoch of the ALMA image used in this work, which is $\sim$6\,\% of its orbital period ($P_b$\,=\,113\,$\pm$\,19\,years). This would be possible from Cycle 12 onward due to spatial resolution considerations following the ALMA observatory long-term configuration schedule\footnote{\url{https://almascience.nrao.edu/observing/observing-configuration-schedule/long-term-configuration-schedule}}. \vspace{0.2cm}

We warn that this exercise is most favorable when performed at similar frequencies (i.e., Band 7). Dust grains with different sizes have different degrees of coupling to the gas and thus different dynamics (e.g., \citealt{2009A&A...493.1125L}; \citealt{2020A&A...642A.224M}), and the extension and center of the emission may change. Indeed, this could explain the shift in the putative Trojan emissions detected in \object{LkCa~15} when comparing Band 6 and 7 data by \citet{2022ApJ...937L...1L}. Hence, to study the future orbital motion of the reported dust emission, it is better to perform observations at similar frequencies to guarantee that we are tracing the same dust population. If the Trojan nature of the $L_{5b}$ emission is confirmed, observations at different frequencies could provide complementary information, which would be valuable to understand the evolution of different dust grain-size populations in the co-orbital regions. These data would be very helpful to test different model predictions investigating the dust population at the two Lagrangian regions. For instance, two independent authors (\citealt{2020A&A...642A.224M}, \citealt{2021A&A...647A.174R}) predict that the $L_4$ region tends to harbor smaller dust grains than $L_5$. \vspace{0.2cm}

Seminal works using these data speculate on the unclear nature of the extended emission associated with planet b (shown in green in the right panel of Fig.~\ref{fig:projection}). \citet{2019ApJ...879L..25I} propose different scenarios: circumplanetary dust, a trace of dust particles trapped in $L_5$, or a jet similar to those detected in accreting protostars. In this work, we identify an additional emission spatially compatible with the potential gravitational well of $L_{5b}$. Hence, it may support the trace of dust from the trailing Lagrangian point hypothesis. In particular, a plausible explanation for this extended emission could be that planet b dominates over the gravity of the forming Trojan and, therefore, it can be stealing the material from the surroundings leading to the starvation of its companion (e.g., \citealt{2009A&A...493.1141C} argue that gas accretion increases the mass difference between the co-orbital pair). \vspace{0.2cm}

It is important to highlight that co-orbital motion does not necessarily mean that the material is orbiting in the surroundings of the minimum of the gravitational wells of the Lagrangian points $L_{4}/L_{5}$. Indeed, the libration amplitude can even enclose both of these points as well as $L_3$ at the same time, what is known as a horseshoe orbit. For this reason, other substructures that are in the orbital path of the planets but do not exactly fall at the minimum of the gravitational wells could also be dust trapped in corotation. In the case of planet b, some substructures are seen coming out from the inner disk and near the location of $L_{4b}$. Nonetheless, this region is so close to the inner disk that it prevents us from reaching any conclusion about its nature. Similarly, halfway between planet c and $L_{4c}$ there is a prominence in the inner edge of the outer disk that falls in the orbital path of the outer planet. Higher angular resolution observations would be required to test if this emission could be separated from the outer disk and whether it moves in corotation with PDS~70\,c. The fact that the most convincing of the signatures compatible with co-orbital dust is at the $L_5$ region rather than $L_4$ might be consistent with the theoretical results found in previous works. For instance, \citealt{2020A&A...642A.224M} point out that planets create an overdensity in $L_5$ compared to $L_4$ that results in an asymmetry in the total bulk mass which is always in favor of the trailing Lagrangian region, and thus it is expected that more massive bodies will be found there than in the leading region.

\section{Conclusions}
\label{sec:concl}

We reanalyzed the combined ALMA observations of PDS~70 presented in \citet{2021ApJ...916L...2B} by performing an independent cleaning to search for dust accumulation in the Lagrangian regions of the two detected protoplanets. We find tentative emission at $\sim$4-$\sigma$ significance at the expected position of the Lagrangian region $L_5$ of planet PDS~70~b, which is thus a candidate precursor for the formation of a co-orbital body, or even the leftovers of a massive Trojan body that has already formed. This would be the third claim of dust trapping in a Lagrangian region of a protoplanet (the former ones were in HD~163296 and LkCa~15 disks). Nonetheless, this is the first time the position of such emission can be associated with the expected location of the Lagrangian region of a confirmed planet. For the sake of confirming or rejecting the Trojan candidate, we propose that future ALMA observations revisit the system. Since the orbit of PDS~70~b is known, we have shown that beyond 2026 we may be able to resolve the co-orbital motion. \vspace{0.2cm}

If confirmed, this work represents observational support to the hypothesis of Trojan bodies being a common consequence of planetary formation. It encourages further surveys to find them in both young and evolved systems. Their existence and properties (chemical and dynamical) would provide additional hints for understanding the evolution of planetary systems as a whole. On the other hand, although ALMA is currently making such discoveries possible, its planed wideband sensitivity upgrade (the top priority initiative for the ALMA2030 Development Roadmap) will be crucial to perform these studies in a much more efficient way.

\begin{acknowledgements}
    We thank the anonymous referee for the helpful comments that improved the manuscript. This work makes use of the following ALMA data: ADS/JAO.ALMA\#2018.A.00030.S., ADS/JAO.ALMA\#2017.A.00006.S, ADS/JAO. ALMA\#2015.1.00888.S. ALMA is a partnership of ESO (representing its member states), NSF (USA), and NINS (Japan), together with NRC (Canada), NSC and ASIAA (Taiwan), and KASI (Republic of Korea), in cooperation with the Republic of Chile. The Joint ALMA Observatory is operated by ESO, AUI/NRAO, and NAOJ. This research has been funded by Spanish MCIN/AEI/10.13039/501100011033 grant PID2019-107061GB-C61, and project No. MDM-2017-0737 Unidad de Excelencia {\em Mar\'{\i}a de Maeztu} - Centro de Astrobiolog\'{\i}a (CSIC-INTA). O.\,B.~-~R. is supported by INTA grant PRE-MDM-07. IdG acknowledges support from grant PID2020-114461GB-I00, funded by MCIN/AEI/10.13039/501100011033. J.\,L.~-~B. acknowledges financial support received from the Ram\'on y Cajal programme (RYC2021-031640-I) funded by  MCIN/AEI/10.13039/501100011033 and the EU “NextGenerationEU”/PRTR.  J.\,L.~-~B. also acknowledges the financial support from "la Caixa" Foundation (ID 100010434) and the European Unions Horizon 2020 research and innovation programme under the Marie Sklodowska-Curie grant agreement No 847648, with fellowship code LCF/BQ/PI20/11760023. 
    A.~R. has been supported by the UK Science and Technology research Council (STFC) via the consolidated grant ST/S000623/1 and by the European Union’s Horizon 2020 research and innovation programme under the Marie Sklodowska-Curie grant agreement No. 823823 (RISE DUSTBUSTERS project).
    M.~B. project has received funding from the European Research Council (ERC) under the European Union’s Horizon 2020 research and innovation programme (PROTOPLANETS, grant agreement No. 101002188). S.F. is funded by the European Union under the European Union’s Horizon Europe Research \& Innovation Programme 101076613 (UNVEIL).
\end{acknowledgements}

\bibliographystyle{aa}
\bibliography{references}

\begin{appendix}

\onecolumn

\section{Gallery of images for different Briggs robust values}

\begin{figure}
  \begin{center}
     \subfigure{\includegraphics[width=130mm]{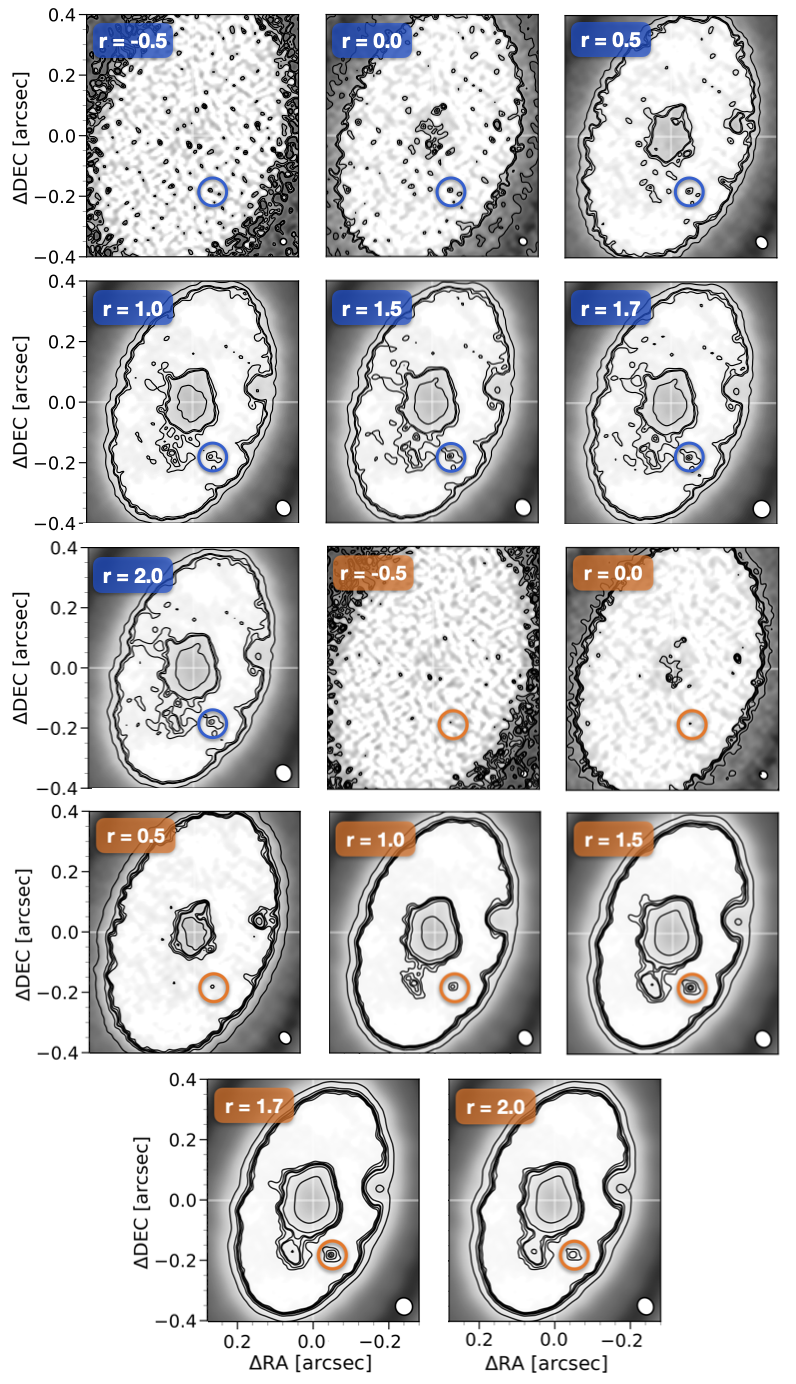}}
  \end{center}
 \caption{Gallery of images with different robust values as indicated in the upper left corner of each panel. The circle indicates the position of the $L_{5b}$ emission. Panels with blue colors correspond with the image without the JvM correction, whose contours are 2, 3, 3.5, and 8-$\sigma$. Panels with orange colors are JvM-corrected images and their contours indicate 3, 4, 5, 5.5, 8, and 22-$\sigma$.}
 \label{fig:set_rob}
\end{figure}

In Fig.\,\ref{fig:set_rob} we show a gallery of images changing the Briggs robust parameter in the CASA cleaning routine. In all of them, the emission at $L_{5b}$ is found with a significance between 3 and 4-$\sigma$, increasing for higher robust parameters since the sensitivity improves. The image shown in this work (see Fig.\,\ref{fig:projection}) uses r = 1.7. This is the value that maximizes the signal-to-noise ratio for our image in the inner cavity and hence provides the highest significance for the tentative detection. In Table\,\ref{tab:rob} we summarize the beam, rms, peak intensity, and significance of the emission for each of those images.

\label{sec:set_rob}

\setlength{\tabcolsep}{10pt}
\begin{table*}[h!]
\centering
\caption[]{Properties for the images with different robust values shown in Fig.\,\ref{fig:set_rob}.}
\label{tab:rob}
\begin{tabular}{cccccc}
\hline
\hline \noalign{\smallskip}
{\bf JvM corrected} & {\bf Briggs Parameter} & {\bf Beam [mas\,$\times$\,mas, $^{\circ}$]} & \makecell{{\bf rms} \\ \bf{[$\mu$Jy\,beam$^{-1}$]}} & \makecell{{\bf Peak intensity} \\ {\bf $L_{5b}$ [$\mu$Jy\,beam$^{-1}$]}} & \makecell{{\bf Significance} \\ {\bf $L_{5b}$ emission}}\\ \noalign{\smallskip}\noalign{\smallskip}
\hline \noalign{\smallskip}
& -0.5 & 21\,$\times$\,21, 29.66 & 25.2 & 79.0 & 3.1-$\sigma$ \\ \noalign{\smallskip} 
& 0.0 & 28\,$\times$\,25, 40.96 & 17.3 & 56.4 & 3.3-$\sigma$ \\ \noalign{\smallskip}
& 0.5 & 44\,$\times$\,36, 51.78 & 12.7 & 41.4 & 3.3-$\sigma$ \\ \noalign{\smallskip}
No & 1.0 & 54\,$\times$\,47, 57.33 & 11.3 & 38.3 & 3.4-$\sigma$ \\ \noalign{\smallskip}
& 1.5 & 58\,$\times$\,51, 58.60 & 10.9 & 39.6 & 3.6-$\sigma$ \\ \noalign{\smallskip}
& 1.7 & 58\,$\times$\,56, 58.71 & 10.9 & 39.9 & 3.7-$\sigma$ \\ \noalign{\smallskip}
& 2.0 & 59\,$\times$\,52, 58.77 & 10.9 & 38.4 & 3.5-$\sigma$ \\ \noalign{\smallskip}
\hline \noalign{\smallskip}
& -0.5 & 21\,$\times$\,21, 29.66 & 18.9 & 59.3 & 3.1-$\sigma$ \\ \noalign{\smallskip} 
& 0.0 & 28\,$\times$\,25, 40.96 & 11.3 & 36.7 & 3.3-$\sigma$ \\ \noalign{\smallskip}
& 0.5 & 44\,$\times$\,36, 51.78 & 7.36 & 25.4 & 3.4-$\sigma$\\ \noalign{\smallskip}
Yes & 1.0 & 54\,$\times$\,47, 57.33 & 3.77 & 16.7 & 3.4-$\sigma$\\ \noalign{\smallskip}
& 1.5 & 58\,$\times$\,51, 58.60 & 3.08 & 18.0 & 5.8-$\sigma$ \\ \noalign{\smallskip}
& 1.7 & 58\,$\times$\,56, 58.71 & 3.22 & 18.5 & 5.7-$\sigma$ \\ \noalign{\smallskip}
& 2.0 & 59\,$\times$\,52, 58.77 & 3.01 & 14.9 & 5.0-$\sigma$ \\ \noalign{\smallskip}
\hline 
\end{tabular}
\end{table*}

\section{Predicted astrometry to detect the co-orbital motion}
\label{sec:astrometry}
\setlength{\tabcolsep}{10pt}
\begin{table*}[h!]
\centering
\caption[]{Coordinates with respect to the center of our image for the peak submillimeter emission of PDS\,70~b and c and their $L_4$ and $L_5$ Lagrangian points. July 2019 is the epoch of the long baseline observation used in the composed image of this work. February 2026 and November 2028 are the minimum predicted dates to measure the co-orbital motion in PDS\,70~b and c orbits, respectively.}
\label{tab:L4L5app}
\begin{tabular}{ccccccc}
\hline
\hline \noalign{\smallskip}
& \multicolumn{2}{c}{\bf July 2019} & \multicolumn{2}{c}{\bf February 2026} & \multicolumn{2}{c}{\bf November 2028} \\ \noalign{\smallskip}\noalign{\smallskip}
\hline \noalign{\smallskip}
& $\Delta$RA {[}mas{]} & $\Delta$DEC {[}mas{]} & $\Delta$RA {[}mas{]} & $\Delta$DEC {[}mas{]} & $\Delta$RA {[}mas{]} & $\Delta$DEC {[}mas{]} \\ \noalign{\smallskip}
\hline \noalign{\smallskip}
PDS~70~b & 70.1 $\pm$ 2.5 & --\,163.0 $\pm$ 3.4 & 100.3 $\pm$ 5.0 & --\,105.4 $\pm$ 7.0 & 106.5 $\pm$ 6.0 & --\,77 $\pm$ 11\\ \noalign{\smallskip} 
$L_{4b}$ &  107.0 $\pm$ 9.0 &  2.8 $\pm$ 9.0 & 90 $\pm$ 10 & 63 $\pm$ 10 & 79 $\pm$ 11 & 87 $\pm$ 10\\ \noalign{\smallskip}
$L_{5b}$ & --\,51.8 $\pm$ 9.0 & --\,174 $\pm$ 15 & --\,0.9 $\pm$ 5.0 & --\,196 $\pm$ 11 & 21.4 $\pm$ 8.0 & --\,194.5 $\pm$ 9.0\\ \noalign{\smallskip}
PDS~70~c & --\,215.1 $\pm$ 1.8 & 37.8 $\pm$ 3.7 & --\,205.2 $\pm$ 4.0 & --\,19.9 $\pm$ 5.0 & --\,199.0 $\pm$ 5.0 & --\,43.7 $\pm$ 6.0 \\ \noalign{\smallskip}
$L_{4c}$ & --\,77 $\pm$ 13 & --\,238 $\pm$ 14 & --\,42 $\pm$ 12 & --\,264 $\pm$ 13 & --\,25 $\pm$ 15 & --\,273 $\pm$ 15\\ \noalign{\smallskip}
$L_{5c}$ & --\,134 $\pm$ 13 &  275 $\pm$ 13 & --\,162 $\pm$ 10 & 251 $\pm$ 14 & --\,173 $\pm$ 11 & 237 $\pm$ 12\\ \noalign{\smallskip}
\hline 
\end{tabular}
\end{table*}

\section{JvM-corrected image}
\label{sec:jvm}
\begin{figure}
  \begin{center}
     \subfigure{\includegraphics[width=95mm]{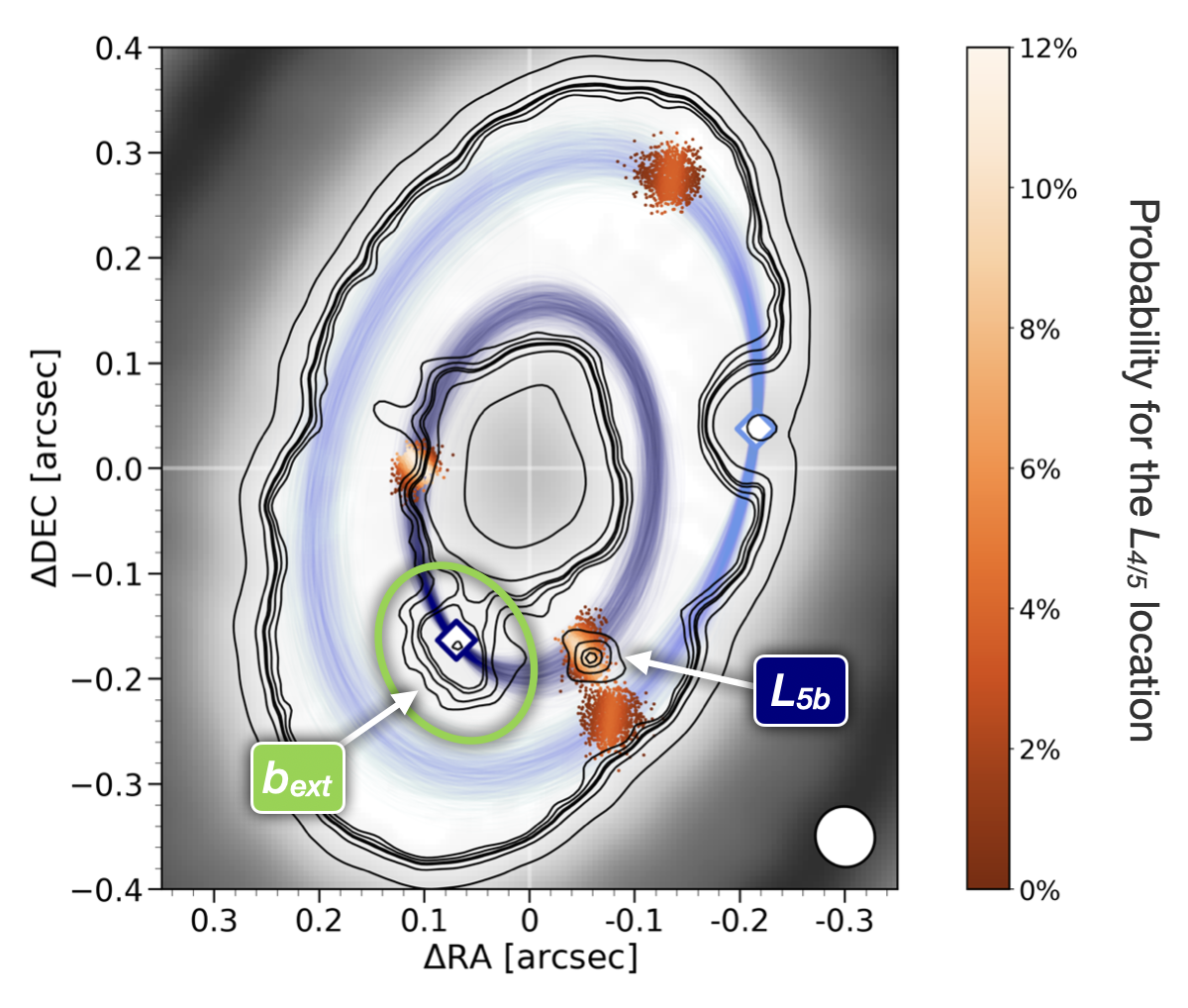}}
  \end{center}
 \caption{Same image as in Fig.\,\ref{fig:projection}, but applying the JvM correction. Contours are 3, 4, 5, 5.5, 8, and 22-$\sigma$. The $L_{5b}$ submillimeter emission here is detected at $\sim$6-$\sigma$ of significance.} 
 \label{fig:jvm}
\end{figure}

In the cleaning process, the final image is the result of combining the model and the residual images. Nonetheless, there is a mismatch between the units of both images (Jy/CLEAN beam and Jy/dirty beam, respectively). Thus, the resulting flux scale is incorrect (\citealt{1995AJ....110.2037J}). Following the methodology of \citet{2021ApJS..257....2C}, we corrected for this JvM effect by rescaling the residual image by the CLEAN beam/dirty beam volumes ratio before combining it with the model. We show the cleaned image in Fig.\,\ref{fig:jvm}. In particular, the emission falling in our interest region $L_{5b}$ turns out to be at a 5.7-$\sigma$ level since the rms when using this correction is 3.2~$\mu$Jy~beam$^{-1}$ and the peak intensity is 18.5~$\mu$Jy~beam$^{-1}$.

\section{Shift between submillimeter and infrared emission of PDS~70\,b}
\label{sec:bshift}

\begin{figure}[h!]
  \begin{center}
     \subfigure{\includegraphics[width=140mm]{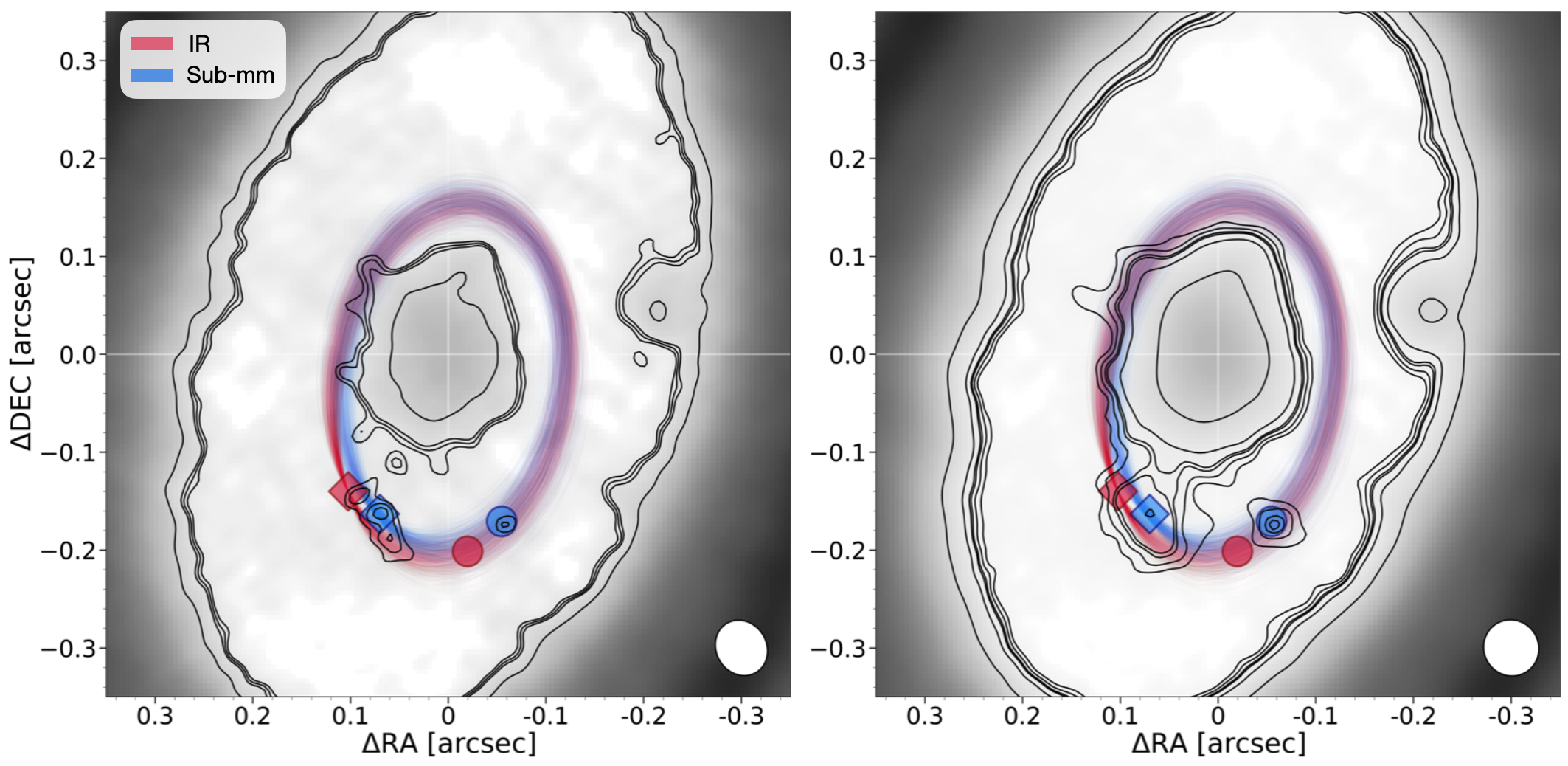}}
  \end{center}
 \caption{Same ALMA image and contours as in Fig.\,\ref{fig:projection} (\textit{left}) and Fig.\,\ref{fig:jvm} (\textit{right}, which is JvM corrected), but including the infrared position of planet PDS~70\,b. Diamond markers indicate the infrared position of PDS~70\,b (in red) and the maximum of the submillimeter emission associated with PDS~70\,b (in blue). The orbits are restricted to those crossing the position of the markers in each case. The big solid circles represent the location of the $L_5$ points associated with the infrared position of the planet (in red) and with the peak of the
 submillimeter emission (in blue), respectively.}
 \label{fig:bshift}
\end{figure}

In order to consider the reported shift between the infrared and submillimeter positions for planet b, we derived the location of the $L_5$ point associated with the infrared position of the planet. We used the position inferred in \citet{2021AJ....161..148W} from the GRAVITY observations in the K band, taken at the same epoch as the ALMA long baseline observations (July 2019): $\Delta$RA = (102.61 $\pm$ 0.09) mas and $\Delta$DEC = (-139.93 $\pm$ 0.24) mas. In Fig.\,\ref{fig:bshift} we compare the orbits of planet b when restricting them to the location of the planet in the infrared (red), and to the maximum of the submillimeter emission as in the main text (blue). The locations of the $L_5$ points for each case are indicated with solid circles. As discussed in the main text, the blue circle perfectly matches the compact emission tentatively detected, while it is $\sim$5$^{\circ}$ in azimuth backward of the $L_5$ point associated with the 
PDS~70\,b infrared position. The orbital paths in both cases (blue and red) cross the detected emission. Hence, this compact emission perfectly lies within the Lagrangian region of the planet, and it might be dust trapped in its potential well, co-orbiting with PDS~70\,b and librating in its $L_5$ point.

\end{appendix}

\end{document}